\documentclass[12pt]{article}

	\usepackage{graphicx,amsmath,amssymb,lineno}
	\usepackage{cite}

	\usepackage{fancyhdr} 

\begin{document}

	\title{A General Formulation\\ for the Stiffness Matrix of Parallel Mechanisms}

	\author{Cyril Quennouelle\thanks{Corresponding author}\\
		\texttt{cquennouelle@gmail.com}
		\and
		Cl\'{e}ment~Gosselin\\
		\texttt{gosselin@gmc.ulaval.ca}\\
		~\\
		Laboratoire de robotique,\\
		D\'{e}partement de g\'{e}nie m\'{e}canique, Universit\'{e} Laval\\
		1065, avenue de la m\'edecine - Qu\'ebec, QC, Canada - G1V 0A6
	}
	\date{}


\maketitle

\selectfont

\begin{abstract}
	{\it 
		Starting from the definition of a stiffness matrix, the authors present a new formulation of the Cartesian stiffness matrix of parallel mechanisms. The proposed formulation is more general than any other stiffness matrix found in the literature since it can take into account the stiffness of the passive joints, it can consider additional compliances in the joints or in the links and it remains valid for large displacements. Then, the validity, the conservative property, the positive definiteness and the relation with other formulations of stiffness matrices are discussed theoretically. Finally, a numerical example is given in order to illustrate the correctness of this matrix.
	}
\end{abstract}

%



\section{Introduction}
	\label{intro}
	A robotic manipulator is a mechanism designed to displace objects in space or in a plane. Therefore, a high precision in the position and orientation of the end-effector and a good repeatability of motion are desirable properties of a manipulator. To fulfil this objective, an accurate model of the mechanism is required. In particular, it is important to be able to precisely characterize the stiffness of the manipulator, i.e., to determine the relation between the loads applied to the mechanism and the resulting displacements. The mathematical object most commonly used to characterize the stiffness of a mechanism is the \emph{stiffness matrix}.

	In the literature, numerous papers deal with the stiffness matrix (SM) of robotic manipulators (See section~\ref{Sec:Litterature}). However, to the best knowledge of the authors, none of them presents a SM that is general and valid for any parallel mechanism (PM), notably, PMs with passive joints that have a non zero stiffness and where some additional compliances (in the joints as well as in the rigid links) are taken into account. The latter correspond to \emph{elastically articulated rigid-body systems} \cite{Angeles} or \emph{compliant mechanisms} \cite{Howellbook} (notably when the compliant joints are modelled using a multi-degree of freedom (DOF) pseudo-rigid body model \cite{su:021008}). Since such a matrix is essential for the quasi-static \cite{quen2009qsm} and the dynamic modelling of these mechanisms, a SM is presented in this paper that considers the external loads, the changes of geometry of the mechanism, the stiffness of actuated and passive joints and even the finite stiffness of the rigid links, for both planar and spatial PMs.

	After an overview of the literature on the SM, the kinematic model of a PM that takes into account the passive joints is recalled. Then, expressions of the potential energy are derived in order to obtain the generalized stiffness matrix (GSM) of a PM and a general and meaningful form of its Cartesian stiffness matrix (CSM).
	The properties of this matrix are then discussed and finally, an application using the CSM is presented in order to illustrate the correctness and the possible applications of the presented matrix.
\section{The Stiffness Matrix in the Literature}
	\label{Sec:Litterature}
	\paragraph*{Definition}
		Usually, a SM is mathematically defined as the Hessian matrix of a potential, i.e., the square matrix of second-order partial derivatives of this potential. For example, the CSM of a planar mechanism is the Hessian of the potential~$\xi_f$ associated to a wrench~$\mathbf{f}$ with respect to the Cartesian coordinates. It is written as
		\begin{equation}
			\mathbf{K}_C=\frac{d^2\xi_f}{d\mathbf{x}_c^2},
			\label{Equ:HessianPoten}
		\end{equation}
		where~$d\mathbf{x}_c$ represents a infinitesimal variation of the pose.	However in many cases, such a potential energy cannot be determined and the latter definition cannot be applied. The SM is then defined as the Jacobian matrix of a wrench. This is written as
		\begin{equation}
			\mathbf{K}_C=\frac{d\mathbf{f}}{d\mathbf{x}_c}.
			\label{Equ:JacobianForce}
		\end{equation}
		It can be noticed that, when the associated potential is known, the conservative wrench is equal to the gradient of~$\xi_f$, and both definitions are equivalent\footnote{In the literature, it is sometimes stated that a wrench is equal to the opposite of the gradient ($\mathbf{f}=-\nabla \xi_f$) and that a stiffness is equal to the opposite of the Jacobian matrix of a wrench ($\mathbf{K}=-\nabla\mathbf{f}$). These definitions lead to the same results.}.

		Surprisingly, the SM of a mechanism submitted to an external load is symmetric only when it is written in a coordinate basis \cite{li2002sca,chen2003sts,AsymHoward,vzefran2002gas,svinin2001sas,ciblak1994acs,zefran1997acc,ciblak1999scs}. It is asymmetric otherwise.
		Chen and Kao add in \cite{CCT1, CCT2, CCT3,GrapStiffMat, chen2005scc}, that a SM is conservative, i.e., the work done by a force resulting from this matrix along a closed path must be equal to zero. Finally, the Hessian matrix of a potential being used to determine the stability of an equilibrium \cite{PolynomialHomotopy,carricato2002cap, svinin2001sas, carricato}, a SM can be either positive-definite (or semi-definite) in a stable equilibrium or not in an unstable position.
	\paragraph*{Literature review}
		In 1980, Salisbury was the first to formulate a SM for serial mechanisms in \cite{Salisbury}. Then, the formula was extended to PMs in which only the stiffness of the actuators was considered \cite{StiffMap, MerletBook}. In fact, both matrices ---which are still often accepted and applied nowadays---, are only valid in very particular conditions, pointed out by Chen, Kao \emph{et al.} in \cite{CCT1, CCT2, CCT3}: they are correct only when the external loads are zero or when the Jacobian matrix of the mechanism is constant. The misconception stems from the improper use of the following equations:
		\begin{equation}
			\left.\begin{array}{rl}
				\delta\mathbf{x}&=\mathbf{J}\delta\boldsymbol{\psi}\\
				\mathbf{f}&=\mathbf{K}_C\delta\mathbf{x}
			\end{array}\right\}
			\label{Equ:Error}
		\end{equation}
		where~$\mathbf{f}$ is the vector of the external loads,~$\delta\boldsymbol{\psi}$ a small displacement of the joints,~$\mathbf{J}$ the Jacobian matrix of the mechanism and~$\delta\mathbf{x}$ a small displacement of the effector in the first equation and a small gap of pose in the second equation. When both equations are used together, a small gap and a small displacement are incorrectly considered as equivalent and the second equation becomes inconsistent: when the external load remains constant, there should be no displacement of the mechanism.
	
		The SM proposed by Chen, Kao \textit{et al.} in \cite{CCT1} is correct for both serial and parallel planar mechanisms and it has been extended to spatial mechanisms in \cite{chen2005scc}. Using screw theory, Griffis and Duffy also noted the influence of an external load on the SM~\cite{griffis1993gsm}. However, the proposed matrices still suffer from some lack of generality: they cannot take into account the stiffness of the passive joints and the degree of mobility (DOM) of the mechanism has to be equal to the DOF of its end-effector platform. This results in a loss of accuracy in the modelling of compliant mechanisms. 
	
		In \cite{zhangthese, PatPassiv}, Zhang and Gosselin studied PMs with a constraining leg whose compliances were modelled as virtual joints. Thus, the SM that they proposed considers the stiffness of some passive joints. However, they did not describe the effects of the external load nor the effect of the internal force. Furthermore, their SM is not formulated in a general way and can only be applied to the type of PMs with a constraining passive leg. Finally, some works have been published that use a SM approaching the one presented in this paper, however without mainly focusing on it. For example, \cite{cho1989das} considers a redundant actuation and \cite{yi2003dae} considers the stiffness of the passive joints .
\section{Model of a Parallel Mechanism}
		\subsection{Geometric Constraint}
			In a PM, the closure of the loops formed by the legs defines $c$~geometrical constraints that have to be satisfied by the joint coordinates. However, since these constraints can be dependent in the case of an overconstrained mechanism, the actual number of independent geometric constraints is~$\mathfrak{C}$ ($\mathfrak{C}\leq c$). The constraints are written as
			\begin{equation}
				\boldsymbol{\mathcal{K}}(\boldsymbol{\theta}) = \mathbf{0}_\mathfrak{C},
				\label{Equ:Contr_GeoGeneral}
			\end{equation}
			where~$\boldsymbol{\theta}$ is the joint coordinate vector of the mechanism, including all joints, actuated and passive. Note that the flexibility of the links and actuators can be taken into account by adding virtual elastic joints in the mechanism~\cite{quennouellethese, PatPassiv, zhangthese}. These additional coordinates are also included in vector~$\boldsymbol{\theta}$.

		\subsection{Generalized Coordinates}
			A vector of generalized coordinates~$\boldsymbol{\psi}$, is defined such that~$\boldsymbol{\lambda}$ the vector of the kinematically dependent coordinates and~$\boldsymbol{\theta}$, the complete joint coordinate vector of the mechanism, always satisfy the geometric constraints. One has:
			\begin{equation}
				\boldsymbol{\lambda}=\boldsymbol{\lambda}(\boldsymbol{\psi})\text{ and }
				\boldsymbol{\theta}=\boldsymbol{\theta}(\boldsymbol{\psi})
				\label{Equ:theta}
			\end{equation}
			where~$\boldsymbol{\lambda}=\left[\lambda_1;\cdots;\lambda_\mathfrak{C}\right]^{T}$ and ~$\boldsymbol{\theta}=\left[\theta_1;\cdots;\theta_m\right]^{T}$ with~$m$ the number of joints in the mechanism and~$\theta_k$ the coordinate associated with the~$k^{\textrm{th}}$ joint. The dimension~$\mathfrak{M}$ of vector~$\boldsymbol{\psi}$ equals the number of DOM of the (kinematically equivalent) mechanism, such that $\mathfrak{M}+\mathfrak{C}=m$.

			\textbf{Structure of vector~$\boldsymbol{\theta}$:} The dependent and the generalized coordinates can be chosen arbitrarily. They can correspond to joint coordinates or to functions of the latter. If joint coordinates are chosen as dependent and generalized coordinates, they can be sorted such that~$\boldsymbol{\theta}$ can be written as~$\boldsymbol{\theta}^T=\left[\boldsymbol{\psi}^T;\boldsymbol{\lambda}^T\right]$.
		\subsection{Kinematic Constraints}			
			The variation of the kinematically dependent joint coordinates is described by a matrix~$\mathbf{G}$ and a matrix~$\mathbf{R}$ defined as
			\begin{equation}
				\mathbf{G}=\frac{d\boldsymbol{\lambda}}{d\boldsymbol{\psi}}
				\text{~and~}
				\mathbf{R}=\frac{d\boldsymbol{\theta}}{d\boldsymbol{\psi}}
					=\begin{bmatrix}\mathbf{1}_\mathfrak{M}\\\mathbf{G}\end{bmatrix}
				\label{Equ:dlambdaGdChiR}
			\end{equation}
			where~$\mathbf{1}_\mathfrak{M}$ stands for the ($\mathfrak{M}\times\mathfrak{M}$) identity matrix.	The above matrices represent the kinematic constraints in a PM.
			The relations between the variation of the joint coordinates and the variation of the generalized coordinates are expressed as 
			\begin{equation}
				d\boldsymbol{\lambda}= \mathbf{G}d\boldsymbol{\psi}\text{ and }d\boldsymbol{\theta}= \mathbf{R} d\boldsymbol{\psi}.
				\label{Equ:dthetaRdChi}
			\end{equation}
		\subsection{Kinematic Model}
			\subsubsection{Pose of the Platform}
				The pose of the platform, represented by a set of parameters~$\mathbf{x}$, is defined as the average pose of the end-effector of all legs of the mechanism. For the~$i^{\textrm{th}}$ leg, the latter is written as~$\mathbf{x}_i^T=\left[\mathbf{c}_i^T;\mathbf{q}_i^T\right]$,~$\mathbf{c}_i$ being the position vector of a chosen point on the platform and~$\mathbf{q}_i$ a set of parameters representing the orientation. The legs of the PM are indexed from~$a$ to~$n$.

				In a planar mechanism,~$\mathbf{q}_i$ is an angle along the~$z$-axis; in a translational spatial mechanism, the orientation is constant thus~$\mathbf{q}_i$ is not used; and in the general 6-DOF spatial case,~$\mathbf{q}_i$ is a quaternion vector describing the orientation of the platform. 
			\subsubsection{Variation of the Pose of the Platform}
				The instantaneous Cartesian variation of the pose of the platform is represented by a vector~$\mathbf{t}$ defined 
				as~$\mathbf{t}^T=\left[\mathbf{v}^T;\boldsymbol{\omega}^T\right]$,~$\mathbf{v}$ being the velocity of a chosen point of the platform and~$\boldsymbol{\omega}$ its angular velocity. 
				In a planar mechanism,~$\mathbf{t}$ is equal to the instantaneous variation of the pose~$\dot{\mathbf{x}}$, i.e.,~$\mathbf{v}=\dot{\mathbf{p}}$ and~~$\boldsymbol{\omega}=\dot{\mathbf{q}}$. But in the spatial case, since the angular velocity~$\boldsymbol{\omega}$ is a 3-coordinate vector, a~$3\times4$ matrix~$\boldsymbol{\Lambda}$ is used to determine~$\boldsymbol{\omega}$ as a function of the variation of the quaternion vector~$\dot{\mathbf{q}}$ (See \cite{angeles2003frm}). Defining a~$6\times7$ matrix~$\mathbf{L}$, the variation of the pose can be written as
				\begin{equation}
					\mathbf{t}=\mathbf{L}\dot{\mathbf{x}},\textrm{~where~}\mathbf{L}=\begin{bmatrix}\mathbf{1}&\mathbf{0}\\\mathbf{0}&\boldsymbol{\Lambda}\end{bmatrix}.
				\end{equation}
				An infinitesimal Cartesian variation of the pose~$d\mathbf{x}_c$ can also be defined as
				\begin{equation}
					d\mathbf{x}_c=\mathbf{t}dt=\mathbf{L}d\mathbf{x},
				\end{equation}
				where~$dt$ represents an infinitesimal period of time. In the  planar case,~$d\mathbf{x}_c=d\mathbf{x}$ but in the spatial case,~$d\mathbf{x}_c$ has~6 components, while~$d\mathbf{x}$ has 7 components.
			\subsubsection{Cartesian Kinematic model}
				The Jacobian matrix~$\mathbf{J}_{\theta}$ of a PM in which all joints ---even the passive ones--- are considered is defined in \cite{quen2009qsm} as the Jacobian matrix of the Cartesian pose with respect to the joint coordinates. It is written as
				\begin{equation}
					\mathbf{J}_{\theta}=\frac{d \mathbf{x}_c}{d \boldsymbol{\theta}}.
					\label{Equ:JacobianComplete}
				\end{equation}
				This matrix is actually composed of the columns of matrices~$\mathbf{J}_{\theta_a}$ to~$\mathbf{J}_{\theta_n}$, the Jacobian matrices of each of the legs considered as an independent serial mechanism. 

				The Jacobian matrix of the pose of the end-effector with respect to the generalized coordinates is noted~$\mathbf{J}$ and is defined as
				\begin{equation}
					\mathbf{J}=\frac{d \mathbf{x}_c}{d \boldsymbol{\psi}}.
					\label{Equ:JacobianGeneralized}
				\end{equation}
			
				The Cartesian kinematic model of the mechanism is written as
				\begin{equation}
					\mathbf{t}=\mathbf{J}_{\theta} \dot{\boldsymbol{\theta}}=\mathbf{J}_{\theta}\mathbf{R} \dot{\boldsymbol{\psi}}
					=\mathbf{J}\dot{\boldsymbol{\psi}}.
					\label{Equ:ModelCine}
				\end{equation}
			\subsubsection{Complete Kinematic Model}
				In a mechanism, especially in a compliant one, the number of DOM~$\mathfrak{M}$ can be larger than the number of degrees of freedom~$\mathfrak{F}$ of the end-effector platform \cite{quennouellethese, su:021008}, thus determining only~$\mathbf{t}$ may not be sufficient to completely determine the configuration of the mechanism. Therefore,~$\mathbf{t}$ comprising~$\mathfrak{F}$ components, ($\mathfrak{M}-\mathfrak{F}$) additional output coordinates are chosen to complete the kinematic model. They are noted~$y_i$ and assembled in a vector~$\mathbf{y}$. These coordinates can be any Cartesian coordinates of others points of the mechanism as well as joint coordinates.
				By assembling all the~$\dot{y}_i$ and the~$\mathfrak{F}$ components of~$\mathbf{t}$ in a vector~$\mathbf{u}$ containing~$\mathfrak{M}$ components, the direct complete kinematic model can be written as
				\begin{equation}
					\mathbf{u}=\begin{bmatrix}\mathbf{t}\\\dot{\mathbf{y}}\end{bmatrix}=
					\begin{bmatrix}\mathbf{J}\\\mathbf{J}_y\\\end{bmatrix}\dot{\boldsymbol{\psi}}=\mathbf{H}\dot{\boldsymbol{\psi}},
					\label{Equ:ModelCineComplet}
				\end{equation}
				where~$\mathbf{H}$ is the~$\mathfrak{M}\times\mathfrak{M}$ complete Jacobian matrix of the mechanism and~$\mathbf{J}_y$ is the~$(\mathfrak{M}-\mathfrak{F})\times\mathfrak{M}$ Jacobian matrix of the~$y_i$ coordinates with respect to the~generalized coordinates.
			\paragraph{Inverse Kinematic Model}
				In a non singular configuration,~$\mathbf{H}$ is invertible. Then, from equation~(\ref{Equ:ModelCineComplet}), the inverse kinematic model of the mechanism is expressed as~: 
				\begin{equation}\dot{\boldsymbol{\psi}}=\mathbf{H}^{-1}\mathbf{u}.
					\label{Equ:ModelCineCompletInverse}
				\end{equation}

				When~$\mathfrak{M}=\mathfrak{F}$, no additional coordinates~$y_i$ are required, thus~$\mathbf{u}=\mathbf{t}$ and~$\mathbf{H}=\mathbf{J}$. In this case, equation~(\ref{Equ:ModelCineCompletInverse}), can be written as
				\begin{equation}
					\dot{\boldsymbol{\psi}}=\mathbf{J}^{-1}\mathbf{t}.
				\end{equation}
			\paragraph{Infinitesimal Variation}
				With the infinitesimal variation of the pose and of the~$y_i$ coordinates, the complete direct and inverse kinematic model are respectively written as
				\begin{equation}
					\begin{bmatrix}d\mathbf{x}_c\\d\mathbf{y}\end{bmatrix}=\mathbf{H}d\boldsymbol{\psi}~~~\textrm{and}~~~d\boldsymbol{\psi}=\mathbf{H}^{-1}\begin{bmatrix}d\mathbf{x}_c\\d\mathbf{y}\end{bmatrix}.
				\end{equation}
\section{Stiffness Matrix of a Parallel Mechanism}
		\subsection{Potential Energy of a Mechanism}
			\subsubsection{Elastic Potential Energy}
			The potential energy stored in the elastic joints of a mechanism, noted~$\xi_\theta$, is written as 
			\begin{equation}
				\xi_\theta=\int_{\boldsymbol{\theta}_0}^{\boldsymbol{\theta}}\boldsymbol{\tau}_\theta^Td\boldsymbol{\theta}=\int_{\boldsymbol{\psi}_0}^{\boldsymbol{\psi}}\boldsymbol{\tau}_\psi^Td\boldsymbol{\psi}+\int_{\boldsymbol{\lambda}_0}^{\boldsymbol{\lambda}}\boldsymbol{\tau}_\lambda^Td\boldsymbol{\lambda}
			\end{equation}
			where~$\boldsymbol{\tau}_j$ is the vector of joint torques/forces associated with the joints corresponding to vector~$\mathbf{j}$ and where~$\boldsymbol{\theta}_0$,~$\boldsymbol{\psi}_0$ and~$\boldsymbol{\lambda}_0$ correspond to the undeformed configurations of the joints. In the particular ---but frequent--- case of elastic joints with constant stiffness, the potential energy is written as
			\begin{equation}
				\xi_\theta=\frac{1}{2}\Delta\boldsymbol{\psi}^T\mathbf{K}_\psi\Delta\boldsymbol{\psi}+\frac{1}{2}\Delta\boldsymbol{\lambda}^T\mathbf{K}_\lambda\Delta\boldsymbol{\lambda},
			\end{equation}
			with~$\Delta\boldsymbol{\psi}=\boldsymbol{\psi}-\boldsymbol{\psi}_0$ and~$\Delta\boldsymbol{\lambda}=\boldsymbol{\lambda}-\boldsymbol{\lambda}_0$ and where~$\mathbf{K}_\psi$ and~$\mathbf{K}_\lambda$ are the (diagonal) joint SMs.
			\subsubsection{Conservative External Load }
			 In a planar mechanism, the potential energy~$\xi_f$ associated to the load~$\mathbf{f}$ applied to the end-effector platform is equal to 
			\begin{equation}
				\xi_f=\int_{\mathbf{x}_0}^{\mathbf{x}}\mathbf{f}^Td\mathbf{x}_c=\int_{\mathbf{x}_0}^{\mathbf{x}}\mathbf{f}^T\mathbf{J}d\boldsymbol{\psi},
				\label{Equ:EnerguXifplan}
			\end{equation}
			where~$\mathbf{x}_0$ corresponds to the unloaded configuration.

			In the spatial case since the angular velocity~$\boldsymbol{\omega}$ and the infinitesimal variation of pose~$d\mathbf{x}_c$ are not integrable, the associated potential cannot be written. However, the instantaneous power of a 6-dimensional external load is defined as
			\begin{equation}
				\dot{\xi_f}=\mathbf{f}_l^T\mathbf{v}+\mathbf{m}^T\boldsymbol{\omega}=\mathbf{f}^T\mathbf{t}=\mathbf{f}^T\mathbf{J}\dot{\boldsymbol{\psi}},
				\label{Equ:EnerguXif}
			\end{equation}
			where~$\mathbf{f}^T=\left[\mathbf{f}_l^T;\mathbf{m}^T\right]$, $\mathbf{f}_l$ representing the 3-dimensional force vector and~$\mathbf{m}$ the 3-dimensional moment vector.
			\subsubsection{Potential Energy of the Mechanism}
			In a planar mechanism, the potential energy~$\xi_f$ due to the external wrench is equal ---apart from a constant~$\xi_0$--- to the energy stored in the mechanism $(\xi_f=\xi_\theta+\xi_0)$. Using eq.(\ref{Equ:dthetaRdChi}) and eq.(\ref{Equ:EnerguXifplan}), this can be written as
			\begin{equation}
				\int_{\boldsymbol{\psi}_{f_0}}^{\boldsymbol{\psi}}\mathbf{f}^T\mathbf{J}d\boldsymbol{\psi}=\int_{\boldsymbol{\psi}_{f_0}}^{\boldsymbol{\psi}}\boldsymbol{\tau}_\psi^Td\boldsymbol{\psi}+\int_{\boldsymbol{\psi}_{f_0}}^{\boldsymbol{\psi}}\boldsymbol{\tau}_\lambda^T\mathbf{G}d\boldsymbol{\psi}+\xi_0,
				\label{Equ:EquationEnergy}
			\end{equation}
			where~$\xi_0$ represents the energy stored in the mechanism in configuration~$\boldsymbol{\psi_{f_0}}$, where~$\mathbf{f}=\mathbf{0}$. This energy is not zero when a preload exists in the compliant joints.
			The infinitesimal variation of eq.(\ref{Equ:EquationEnergy}) is also valid for a spatial mechanism. It is written as
			\begin{equation}
				\mathbf{f}^T\mathbf{J}d{\boldsymbol{\psi}}=\boldsymbol{\tau}_\psi^Td{\boldsymbol{\psi}}+\boldsymbol{\tau}_\lambda^T\mathbf{G}d{\boldsymbol{\psi}}.
				\label{Equ:EquationEnergyInstant}
			\end{equation}
		\subsection{Static Equilibrium}
			Differentiating eq.(\ref{Equ:EquationEnergy}) with respect to the generalized coordinates~$\boldsymbol{\psi}$ leads to the generalized static equilibrium of a mechanism subjected to an external wrench, which is written as
			\begin{equation}
				\frac{d\xi_f}{d\boldsymbol{\psi}}=\frac{d\xi_\theta}{d\boldsymbol{\psi}}+\frac{d\xi_0}{d\boldsymbol{\psi}}\Leftrightarrow\mathbf{J}^T\mathbf{f}=\boldsymbol{\tau}_\psi+\mathbf{G}^T\boldsymbol{\tau}_\lambda.
				\label{Equ:derivEnergy}
			\end{equation}
			The right-hand side of the latter relation is also valid in the spatial case, since it corresponds to the differentiation of eq.(\ref{Equ:EquationEnergyInstant}) with respect to~$d{\boldsymbol{\psi}}$. Introducing the generalized force~$\boldsymbol{\tau}_M$, eq.(\ref{Equ:derivEnergy}) is equivalent to
			\begin{equation}
				\boldsymbol{\tau}_M=\boldsymbol{\tau}_\psi+\mathbf{G}^T\boldsymbol{\tau}_\lambda-\mathbf{J}^T\mathbf{f}=\mathbf{0}.
				\label{Equ:Sequilibre}
			\end{equation}
			Note that in the most general case, the stiffness of the joints is not constant and the corresponding forces/torques are defined as
			\begin{equation}
				\medskip\boldsymbol{\tau}_\psi=\displaystyle\int_{\boldsymbol{\psi}_0}^{\boldsymbol{\psi}}\mathbf{K}_\psi d\boldsymbol{\psi}
				\textrm{~and~}
				\boldsymbol{\tau}_\lambda=\displaystyle\int_{\boldsymbol{\lambda}_0}^{\boldsymbol{\lambda}}\mathbf{K}_\lambda d\boldsymbol{\lambda}.
			\label{Equ:Taus}
		\end{equation}
	\subsection{Generalized Stiffness Matrix}
		The GSM~$\mathbf{K}_M$ of a mechanism is defined as the Hessian matrix of the potential energy with respect to the generalized coordinates. However when no expression for the potential energy is known ---such as in the case of a spatial mechanism--- the equivalent following definition is used:~$\mathbf{K}_M$ is equal to the differentiation of the generalized force~$\boldsymbol{\tau}_M$ with respect to~$\boldsymbol{\psi}$. 
		Therefore, using eqs.~(\ref{Equ:Sequilibre}) and (\ref{Equ:Taus}), it is obvious that~$\mathbf{K}_M$ is not constant and depends on the stiffness of the joints and the geometric configuration of the mechanism. 
		\begin{equation}
			\begin{aligned}
				\dfrac{d\boldsymbol{\tau}_M}{d\boldsymbol{\psi}}=\frac{d}{d\boldsymbol{\psi}}&\left(\int_{\boldsymbol{\psi}_0}^{\boldsymbol{\psi}}\mathbf{K}_{\psi}d\boldsymbol{\psi}\right.\\
				\medskip&\left.+\mathbf{G}^T\int_{\boldsymbol{\psi}_0}^{\boldsymbol{\psi}}\mathbf{K}_\lambda\mathbf{G}d\boldsymbol{\psi}-\mathbf{J}^T\mathbf{f}\right),
			\end{aligned}
		\end{equation}
		which leads to 
		\begin{equation}
			\frac{d\boldsymbol{\tau}_M}{d\boldsymbol{\psi}}=\mathbf{K}_\psi+\mathbf{K}_I+\mathbf{K}_E,
			\label{Equ:GenStiff001}
		\end{equation}
		where
		\begin{equation}
			\left\{\begin{aligned}
				\mathbf{K}_I&=\frac{d}{d\boldsymbol{\psi}}\left(\mathbf{G}^T\int_{\boldsymbol{\psi}_0}^{\boldsymbol{\psi}}\mathbf{K}_\lambda\mathbf{G}d\boldsymbol{\psi}\right)\\
				\mathbf{K}_E&=\frac{d}{d\boldsymbol{\psi}}\left(-\mathbf{J}^T\mathbf{f}\right)
			\end{aligned}\right.
			\label{Equ:ElementAB}
		\end{equation}
		Detailed expressions are derived for~$\mathbf{K}_I$ and~$\mathbf{K}_E$ in the next subsections.
		\subsubsection{Matrix~$\mathbf{K}_E$}
			\label{Sec:ElementB}
			The impact of the external wrench on the configuration of the mechanism is governed by the equation~$\boldsymbol{\tau}_f=\mathbf{J}^T\mathbf{f}$, where~$\boldsymbol{\tau}_f$ is the vector of joint force/torque due to the external wrench~$\mathbf{f}$. In this paper, the external load~$\mathbf{f}$ is assumed to be independent from the configuration, thus~$d\mathbf{f}/d\boldsymbol{\psi}=\mathbf{0}$. Matrix~$\mathbf{K}_E$ is equal to 
			\begin{equation}
				\mathbf{K}_E=\frac{d}{d\boldsymbol{\psi}}\left(-\mathbf{J}^T\mathbf{f}\right)=-\frac{d\mathbf{J}^T}{d\boldsymbol{\psi}}\mathbf{f}.
				\label{Equ:MatKE}
			\end{equation}
			The derivative of the Jacobian matrix~$d\mathbf{J}^{T}/d\boldsymbol{\psi}$ is a tensor of order~$3$. Although it is not a commonly used mathematical object, its manipulation presents no particular difficulty (See \cite{CCT1}). In practice, one can differentiate~$\mathbf{J}^{T}\mathbf{f}$ considering~$\mathbf{f}$ as a constant wrench.
			This matrix captures the effect of a change of geometry on~$\boldsymbol{\tau}_f$ and therefore on~$\boldsymbol{\tau}_M$. 
			Matrix~$\mathbf{K}_E$ is written as
			\begin{equation}
				\mathbf{K}_E=-\left[(\frac{d\mathbf{J}^{T}}{d\psi_1}\mathbf{f});\cdots;(\frac{d\mathbf{J}^{T}}{d\psi_\mathfrak{M}}\mathbf{f})\right],
			\end{equation}
			where~$\psi_i$ is the~$i^{\textrm{th}}$ joint coordinate of~$\boldsymbol{\psi}$ and~$(d\mathbf{J}^{T}/d\psi_i)\mathbf{f}$ is a vector forming the~$i^{\textrm{th}}$ column of~$\mathfrak{M}\times\mathfrak{M}$ matrix~$\mathbf{K}_E$. It can be noted that matrix~$\mathbf{K}_E$ is indeed equal to the opposite of the matrix noted~$\mathbf{K}_G$, the active SM introduced in~\cite{CCT1,CCT2,CCT3,chen2005scc,li2002sca,chen2003sts}.
		\subsubsection{Matrix~$\mathbf{K}_I$}
			Developing eq.(\ref{Equ:ElementAB}), matrix~$\mathbf{K}_I$ is composed of two elements:
			\begin{equation}
				\mathbf{K}_I=\frac{d\mathbf{G}^{T}}{d\boldsymbol{\psi}}\boldsymbol{\tau}_\lambda+\mathbf{G}^T\mathbf{K}_\lambda\mathbf{G}.
				\label{Equ:MatKI}
			\end{equation}
			Similarly to matrix~$\mathbf{K}_E$, matrix~$\mathbf{K}_I$ contains a tensor of order~$3$, namely $(d\mathbf{G}^T/d\boldsymbol{\psi})$. 
			Therefore, a matrix~$\mathbf{K}_{IG}$ that captures the effect of the change of geometry of the kinematic constraints, is defined as 
			\begin{equation}
				\mathbf{K}_{IG}=\frac{d\mathbf{G}^{T}}{d\boldsymbol{\psi}}\boldsymbol{\tau}_\lambda=\left[(\frac{d\mathbf{G}^{T}}{d\psi_1}\boldsymbol{\tau}_\lambda)~\cdots~(\frac{d\mathbf{G}^{T}}{d\psi_\mathfrak{M}}\boldsymbol{\tau}_\lambda)\right]
				\label{Equ:KR}
			\end{equation}
			where~$(d\mathbf{G}^{T}/d\psi_i)\boldsymbol{\tau}_\lambda$ is a vector forming the~$i^{\textrm{th}}$ column of~$\mathfrak{M}\times\mathfrak{M}$ matrix~$\mathbf{K}_{IG}$. Recalling the definition of matrix~$\mathbf{G}$ (eq.(\ref{Equ:dlambdaGdChiR})), matrix~$\mathbf{K}_{IG}$ can also be defined as
			$\mathbf{K}_{IG}=(d^2\boldsymbol{\lambda}^{T}/d\boldsymbol{\psi}^2)\boldsymbol{\tau}_\lambda$.
			Matrices~$\mathbf{K}_{IG}$ and~$\mathbf{G}^T\mathbf{K}_\lambda\mathbf{G}$ are functions of the generalized coordinates and represent the contribution of the kinematically constrained joints to the stiffness of the mechanism. This contribution is assembled in matrix~$\mathbf{K}_I$.
		\subsubsection{Generalized Stiffness Matrix}
			Finally, combining eq.(\ref{Equ:GenStiff001}), eq.(\ref{Equ:MatKE}) and eq.(\ref{Equ:MatKI}), the stiffness of the mechanisms is described in the domain of the generalized coordinates, by matrix~$\mathbf{K}_M$ which is written as
				\begin{equation}
					\mathbf{K}_M=\mathbf{K}_\psi+\mathbf{K}_I+\mathbf{K}_E.
					\label{Equ:Stiffmechanism}
				\end{equation}
			This matrix includes the three contributions that determine the stiffness of a mechanism, namely: the stiffness of the kinematically unconstrained joints ($\mathbf{K}_\psi$), the stiffness due to the dependent coordinates (passive joints and additional compliances) and the internal torques/forces ($\mathbf{K}_I$), and the stiffness due to the external loads ($\mathbf{K}_E$). Note that gravity can also easily be taken into account as additional external forces applied at different point of the mechanism.
	\subsection{Cartesian Stiffness Matrix}
		The definition of the CSM as~$d\mathbf{f}_m/d\mathbf{x}_c$ or~$-d\mathbf{f}/d\mathbf{x}_c$ is valid for planar and spatial mechanisms. Using the chain rule, the following derivation can be performed:
		\begin{equation}
			\mathbf{K}_C=-\left(\frac{d\mathbf{x}_c}{d\mathbf{f}}\right)^{-1}=-\left(\frac{d\mathbf{x}_c}{d\boldsymbol{\psi}}\frac{d\boldsymbol{\psi}}{d\boldsymbol{\tau}_M}\frac{d\boldsymbol{\tau}_M}{d\mathbf{f}}\right)^{-1},
		\end{equation}
		where~$d\mathbf{x}_c/d\boldsymbol{\psi}$ is the~$\mathfrak{F}\times\mathfrak{M}$ Jacobian matrix~$\mathbf{J}$ defined in eq.(\ref{Equ:JacobianGeneralized}); matrix~$d\boldsymbol{\psi}/d\boldsymbol{\tau}_M$ exists since it is the~$\mathfrak{M}\times\mathfrak{M}$ generalized compliance matrix and it is equal to the inverse of~$\mathbf{K}_M$ defined in eq.(\ref{Equ:Stiffmechanism}); finally using eq.(\ref{Equ:Sequilibre}), ~$d\boldsymbol{\tau}_M/d\mathbf{f}$ is a~$\mathfrak{M}\times\mathfrak{F}$ matrix equal to~$-\mathbf{J}^T$. Thus, the CSM is a~$\mathfrak{F}\times\mathfrak{F}$ matrix equal to
		\begin{equation}
			\mathbf{K}_C=\left(\mathbf{J}\left(\mathbf{K}_\psi+\mathbf{K}_I+\mathbf{K}_E\right)^{-1}\mathbf{J}^T\right)^{-1}.
			\label{Equ:CartStiff}
		\end{equation}

		When~$\mathfrak{M}=\mathfrak{F}$ and when~$\mathbf{J}$ is not singular, the relationship between the stiffness in the generalized domain and in the Cartesian domain can be written under a familiar form, namely
		\begin{equation}
			\mathbf{K}_C=\mathbf{J}^{-T}\mathbf{K}_M\mathbf{J}^{-1}.
			\label{Equ:CartStiffgenStiff}
		\end{equation}
		And the inverse relation is written as
		\begin{equation}
			\mathbf{K}_M=\mathbf{J}^{T}\mathbf{K}_C\mathbf{J}.
		\end{equation}
		\paragraph{Complete Stiffness Matrix}Using the complete kinematic model (eq.(\ref{Equ:ModelCineComplet}) and eq.(\ref{Equ:ModelCineCompletInverse})), the complete SM ---a~$\mathfrak{M}\times\mathfrak{M}$ matrix noted~$\mathbf{K}_U$--- can be written as
		\begin{equation}
			\mathbf{K}_U=\mathbf{H}^{-T}\mathbf{K}_M\mathbf{H}^{-1}.
			\label{Equ:CompleteStiffnessMatrix}
		\end{equation}
\section{Properties of the Matrix}	
	\subsection{Symmetry}
		\label{Sec:Conservativity}
		The properties of the SM for mechanisms without stiff passive joints nor additional compliance has been intensively discussed in the literature~\cite{AsymHoward,chen2005scc,chen2003sts,chen2000gmm,carricato,svinin2001sas,zefran1997acc,ciblak1994acs,ciblak1999scs}. This matrix, noted~$\mathbf{K}_C^0$, is written as
		\begin{equation}
			\mathbf{K}_C^0=\mathbf{J}^{-T}\left(\mathbf{K}_\psi+\mathbf{K}_E\right)\mathbf{J}^{-1}.
		\end{equation}
		In matrix~$\mathbf{K}_C^0$, matrix~$\mathbf{K}_\psi$ is symmetric by definition and matrix~$\mathbf{K}_E$ is symmetric only when it is expressed in a coordinate basis, i.e., a basis satisfying Schwarz's theorem. 
		For example in a~2-DOF planar mechanism, matrix~$\mathbf{K}_E$ is symmetric when the Cartesian coordinates~$(x,y)$ are used and non-symmetric when the polar coordinates~$(r,\vartheta)$ are used \cite{li2002sca}. In a spatial mechanism, since no coordinate basis can be used to describe a~6-DOF mechanism, matrix~$\mathbf{K}_E$ is not symmetric. 
		Moreover, even if the CSM~$\mathbf{K}_C^0$ can be asymmetric, it is conservative~\cite{CCT2,GrapStiffMat,CCT3,chen2005scc,CCT1}.

		The GSM~$\mathbf{K}_M$ comprises one additional term when the passive joints or the links are compliant, namely~$\mathbf{K}_I$. Since this SM~$\mathbf{K}_I$ has been calculated as the Hessian matrix of~$\xi_\lambda$, the elastic potential energy stored in joints~$\boldsymbol{\lambda}$ with respect to the generalized coordinates~$\boldsymbol{\psi}$ that form a coordinate basis,~$\mathbf{K}_I$ is symmetric and conservative. 
		Thus,~$\mathbf{K}_M$, which corresponds to the sum of~$\mathbf{K}_\psi$,~$\mathbf{K}_E$ and~$\mathbf{K}_I$ has the same symmetric and conservative properties as~$\mathbf{K}_E$. Hence, the fact that~$\mathbf{J}$ is square or not (when additional compliances are added in the mechanism) has no influence on the symmetry and the conservativeness of~$\mathbf{K}_C$. Therefore, matrix~$\mathbf{K}_C$ defined in eq.(\ref{Equ:CartStiff}) has the same symmetric and conservative properties as matrix~$\mathbf{K}_C^0$.
	\subsection{Positive Definite Property}	
		In \cite{CCT1}, the authors show that the SM of a mechanism \emph{without} compliant passive joints can be positive definite, positive semi-definite or non-positive definite depending on the configuration and the external forces. Similarly, the SM presented in this paper can be positive definite, positive semi-definite or non-positive definite. Indeed, a SM is by definition, a measure of the stability of an equilibrium and a positive definite matrix is required to maintain stability.
 		\begin{figure}[t]
			\centering
			\includegraphics{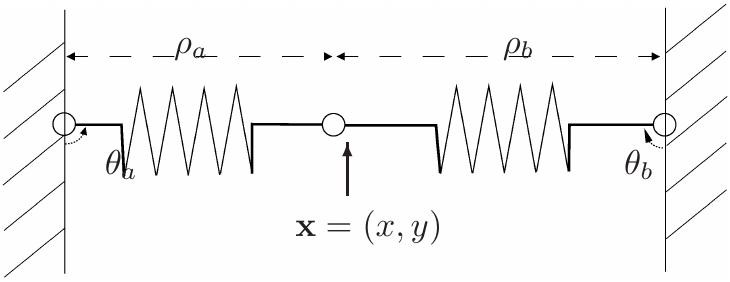}
			\caption{2-DOF parallel mechanism in an unstable static equilibrium.}
			\label{Fig:EquInstable}
		\end{figure}
		For example, one can compute the SM of the 2-DOF mechanism shown in Fig.\ref{Fig:EquInstable}. When both springs are in tension ($\rho_a>\rho_{a_0}$ and~$\rho_b>\rho_{b_0}$), the mechanism is in a stable static equilibrium and the SM is positive-definite. When the springs are in compression, the mechanism is in an unstable static equilibrium and the SM is non-positive definite since the eigenvalue of the matrix corresponding to the vertical axis is negative. Moreover, it can be noticed that by choosing~$\theta_a$ and~$\rho_a$ as generalized coordinates, the configuration of the mechanism shown in Fig.\ref{Fig:EquInstable} is not kinematically singular (i.e.,~$\det\mathbf{J}\neq0$ with~$\boldsymbol{\psi}=[\theta_a,\rho_a]$).

		Any other proposed CSM that does not take into account neither the stiffness of the passive joints nor the effects of the changes of geometry (through matrices~$\mathbf{K}_I$ and~$\mathbf{K}_E$) will not allow the description of this phenomenon of instability.
	\subsection{Other Stiffness Matrices}
		The CSMs found in the literature can be easily obtained from the matrix presented here, since the latter is more general:
		\begin{itemize}
 			\item[$\bullet$] In the literature, the DOM~$\mathfrak{M}$ is almost always equal to~$\mathfrak{F}$ the degree of freedom of the end-effector platform, thus~$\mathbf{J}^{-1}$ exists when the mechanism is not in a singular configuration. Therefore, the comparisons in this subsection can be made with eq.(\ref{Equ:CartStiffgenStiff}).
 			\item[$\bullet$] {The matrices for serial mechanisms (Salisbury \cite{Salisbury}, Chen and Kao \cite{CCT1})} in which there are no passive joints, i.e.,~$\boldsymbol{\theta}=\boldsymbol{\psi}$. Thus there are no internal wrenches and~$\mathbf{K}_I=\mathbf{0}$.
 			\item[$\bullet$] {The matrices when the external wrench~$\mathbf{f}$ is zero or the Jacobian matrix is constant (Salisbury \cite{Salisbury})}. Both cases give~$\mathbf{K}_E=\mathbf{0}$.
			\item[$\bullet$] {The ``\emph{infinite}'' SM of conventional mechanisms} that are considered as not sensitive to external wrenches. In these cases, the stiffness of the actuators is considered infinite and that of the passive joints equal to~$0$, therefore eq.(\ref{Equ:Stiffmechanism}) gives~$\mathbf{K}_M=\text{diag}(\infty)$ and~eq.(\ref{Equ:CartStiff}) gives~$\mathbf{K}_C=\infty_{3\times 3}$.
			\end{itemize}
	\subsection{Use of a Stiffness Matrix}
		In the literature on the \emph{theory of mechanisms}, the research papers mainly focus on the~$\mathfrak{F}\times\mathfrak{F}$ CSM. However, this matrix is not the most useful SM to describe the behaviour of a PM.

		First, the GSM~$\mathbf{K}_M$ is simpler to obtain and allows a complete description of the mechanism, notably when~$\mathfrak{M}>\mathfrak{F}$,  and of the relation between wrenches and displacements ---so can the~$\mathfrak{M}\times\mathfrak{M}$ CSM~$\mathbf{K}_U$ but the latter is more expensive to compute. Note that the concept of GSM is recent, because before the understanding of the influence of external wrenches on the stiffness in the 1990's,~$\mathbf{K}_M$ could not be distinguished from~$\mathbf{K}_\psi$.

		Then, more important than the changes of coordinate basis that, \emph{in fine}, correspond to the choice between Cartesian or GSM, the idea of characterizing a mechanism by a \emph{stiffness} matrix is not very relevant. Actually, this choice seems to be due to a mimetism with springs that are generally characterized by their stiffness. In practice, the computation of the SM is not as useful as that of the \emph{compliance matrix} that determines the displacement of the mechanism due to a variation of the wrenches applied on it. For example, the computation of the quasi-static model of a compliant PM~\cite{quen2009qsm} requires the determination of the \textbf{generalized compliance matrix}, noted~$\mathbf{C}_M$ and equal to~$\mathbf{K}_M^{-1}$.	The relations between these matrices are written as
			\begin{equation}
					\mathbf{C}_U=\mathbf{H}\mathbf{C}_M\mathbf{H}^T=\begin{bmatrix}\medskip\mathbf{C}_C&\mathbf{J}~\mathbf{C}_M\mathbf{J}_y^T\\\mathbf{J}_y\mathbf{C}_M\mathbf{J}^T&\mathbf{J}_y~\mathbf{C}_M\mathbf{J}_y^T\end{bmatrix},
			\end{equation}
		where~$\mathbf{C}_C=\mathbf{J}\mathbf{C}_M\mathbf{J}^T$ is the~$\mathfrak{F}\times\mathfrak{F}$ Cartesian compliance matrix and~$\mathbf{C}_U$, the~$\mathfrak{M}\times\mathfrak{M}$ Cartesian compliance matrix.
	\subsection{Alternative Formulation}
		Following the definition of matrices~$\mathbf{K}_E$ and~$\mathbf{K}_{IG}$, the calculation of the GSM~$\mathbf{K}_M$ (eq.(\ref{Equ:Stiffmechanism})) requires the differentiation of matrices~$\mathbf{J}$ and~$\mathbf{G}$ with respect to the generalized coordinates~$\boldsymbol{\psi}$. Yet, in practice an analytical expression of these matrices as functions of~$\boldsymbol{\psi}$ is not always known and thus, their differentiation might not be performed simply. For this reason, another formulation of~$\mathbf{K}_M$ has been developed that only requires differentiation with respect to $\boldsymbol{\theta}$. This alternative formulation is detailed in appendix~\ref{Sec:AlterKchi} and is used in the application.
		
\section{Application to a Compliant 3-RPR Mechanism}
	In this section, the stiffness of a compliant~3-RPR mechanism presented in \cite{PolynomialHomotopy} is studied. This example is relatively simple and the compliant joints are modelled as 1-DOF joints in order to obtain short and simple formal equations.
	First, the details of the modelling are given, then a comparison between the different SMs proposed in the literature is performed and finally one new possibility offered by the presented SM is used to show the impact of the stiffness of the passive joints on the behaviour of this mechanism.
		\begin{figure}[t]
			\centering
			\includegraphics[height=5cm]{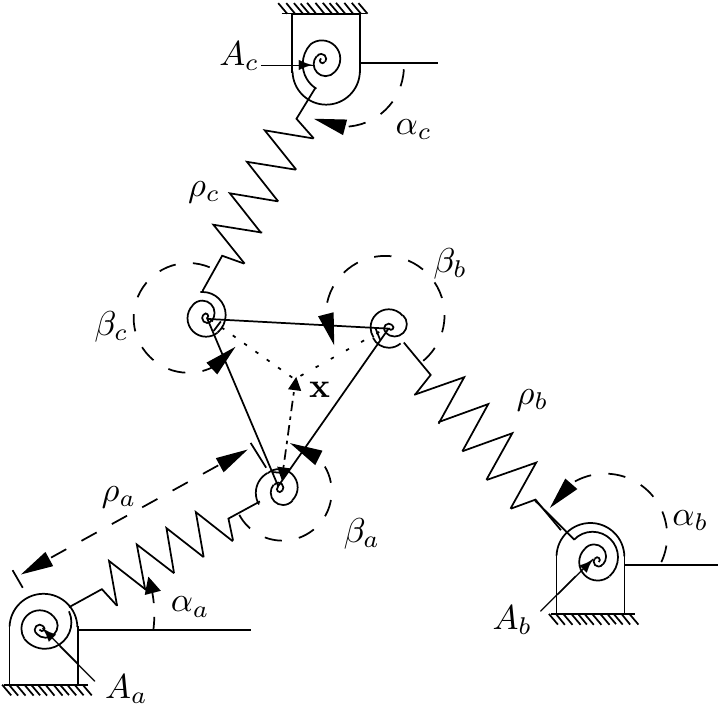}
			\caption{3-RPR planar compliant mechanism.}
			\label{Fig:3PRP}
		\end{figure}	
	\subsection{Modelling of the Mechanism}
		\subsubsection{Geometry of the Mechanism}
			\paragraph{Geometric Parameters of the Legs}
				Each leg~$i$, indexed from $a$ to $c$, is defined by the following parameters:\\
				$\bullet$ All elastic joints are modelled as 1-DOF joints, thus the DOM of the mechanism is equal to the degree of freedom of the platform:~~$\mathfrak{M}=\mathfrak{F}=3$.\\
				$\bullet$ The angles associated with the first revolute joints of each leg are noted~$\alpha_i$. Their unloaded configurations are:~$\alpha_{a0}=0.5404$\,rad,~$\alpha_{b0}=2.0695$\,rad and~$\alpha_{c0}=-1.8252$\,rad.\\
				$\bullet$ The coordinates of the prismatic joints are noted~$\rho_i$. Their unloaded configurations are~$\rho_{a0}=583.10$\,mm,~$\rho_{b0}=683.22$\,mm and~$\rho_{c0}=688.18$\,mm.\\
				$\bullet$ The angles associated with the second revolute joints of each leg are noted~$\beta_i$. Their unloaded configurations are:~$\beta_{a0}=1.0304$\,rad,~$\beta_{b0}=-4.6875$\,rad and~$\beta_{c0}=1.3016$\,rad.\\
				$\bullet$ The position of the points of the base are~$A_a=(x_{a0}, y_{a0})=(-50,-50)$\,cm, $A_b=(x_{b0},y_{b0})$ $=(50,-50)$\,cm and~$A_c=(x_{c0}, y_{c0})=(-50,76)$\,cm.\\
				$\bullet$ The distance between all second revolute joints and the effector's point of reference:~$l_a=l_b=l_c=l$.
			\paragraph{Pose of the Platform}
				The pose of the platform, when considered the end-effector of the~$i^{\textrm{th}}$ leg, is written as
				\begin{equation}
					\mathbf{x}_i=\begin{bmatrix}x_i\\y_i\\\phi_i
					\end{bmatrix}=\begin{bmatrix}
						x_{i0}+\rho_ic\alpha_i+l_ic{(\alpha_i+\beta_i)}\\
						y_{i0}+\rho_is\alpha_i+l_is{(\alpha_i+\beta_i)}\\
						\alpha_i+\beta_i\\
					\end{bmatrix},
				\end{equation}
				where $c$ stands for $\cos$ and $s$ for $\sin$. The pose is defined such that~$\mathbf{x}_0=\mathbf{0}$ when the external forces/torques are~$\mathbf{f}=\mathbf{0}$.
			\paragraph{Geometric Constraints on the Platform}
				The platform of this PM is a rigid body. Hence, the distance between each of the attachment points of the legs on the platform must remain constant. 
				The position of the attachment point of leg~$i$ is noted~$\mathbf{C}_i$ and is written as
				\begin{equation}
					\mathbf{C}_i=\left[x_{i0}+\rho_ic\alpha_i;~ y_{i_0}+\rho_is\alpha_i\right]^T.
				\end{equation}
				The distance~$\overline{C_iC_j}$ between 2 points~$C_i$ and~$C_j$ can be calculated with the following equation,
				\begin{equation}
					\begin{aligned}
					\overline{C_iC_j}&=\sqrt{(\mathbf{C}_j-\mathbf{C}_i)^T(\mathbf{C}_j-\mathbf{C}_i)}\\&=\sqrt{(x_{Cj}-x_{Ci})^2+(y_{Cj}-y_{Ci})^2}=L_{ij}
					\end{aligned}
					\label{Equ:Plateformsidelength}
				\end{equation}
				where~$L_{ij}$ is the constant distance between~$C_i$ and~$C_j$.
				These constraint equations can then be written as 
				\begin{equation}
					Q_{ij}=\overline{C_iC_j}^2-L_{ij}^2,~ij\in\left\{a,b,c\right\}^2,~i\neq j.
				\end{equation}
			\paragraph{Geometric Constraints and Generalized Coordinates}
				Since there are two independent kinematic loops in this planar mechanism, 6 constraints have to be satisfied. This mechanism has 9 joints and thus its number of DOM is 3.

				The 3 coordinates~$\rho_i$ are arbitrarily chosen as the generalized coordinates. Thus, the vector of generalized coordinates is written as~$\boldsymbol{\psi}=\left[\rho_a;\rho_b;\rho_c\right]^T$ and the vector of all the joint coordinates in the mechanism is written as
				\begin{equation}
					\boldsymbol{\theta}=\begin{bmatrix}\boldsymbol{\psi}\\\boldsymbol{\lambda}\end{bmatrix}=\begin{bmatrix}\left[\rho_a;\rho_b;\rho_c\right]^T\\		\left[\alpha_a;\beta_a;\alpha_b;\beta_b;\alpha_c;\beta_c\right]^T\end{bmatrix},
				\end{equation}
				where~$\boldsymbol{\lambda}=\left[\alpha_a;\beta_a;\alpha_b;\beta_b;\alpha_c;\beta_c\right]^T$ is the vector of the dependent joint coordinates.

				The rigidity of the platform must always be satisfied, i.e., the position ($x_i,y_i$) of the end of the 3 legs must be equal\footnote{The third component of the pose, representing the orientation of the platform is not used because this orientation is not a function of~$\boldsymbol{\psi}$, the generalized coordinates.} and the distance between the attachment points must always remain constant. Thus, the constraint function for a kinematic loop is written as :
				\begin{equation}
					\boldsymbol{\mathcal{K}}_{ij}(\boldsymbol{\theta})=\begin{bmatrix}x_i-x_j\\y_i-y_j\\Q_{ij}\end{bmatrix}=\mathbf{0}.
					\label{Equ:ConstraintsCij}
				\end{equation}
				And the vector of the kinematic constraints for the whole mechanism is defined as
				\begin{equation}
					\boldsymbol{\mathcal{K}}(\boldsymbol{\theta})=\begin{bmatrix}
						\boldsymbol{\mathcal{K}}_{ab}(\boldsymbol{\theta})\\
						\boldsymbol{\mathcal{K}}_{ac}(\boldsymbol{\theta})
					\end{bmatrix}.
					\label{Equ:ConstraintsC}
				\end{equation}
		\subsubsection{Kinematics: Infinitesimal Variations}
			\paragraph{Rigidity of the Platform}
				The differentiation of the square of the distance between the attachment points on the platform with respect to the joint coordinates is calculated as
				\begin{equation}
					\left\{\begin{aligned}
						\frac{dQ_{ij}}{d\rho_i}=
						-2&c\alpha_i(x_{j0}+\rho_jc\alpha_j-x_{i0}-\rho_ic\alpha_i)\\
						+2&s\alpha_i(y_{j0}+\rho_js\alpha_j-y_{i0}-\rho_is\alpha_i),\\
						\frac{dQ_{ij}}{d\alpha_i}=
						~~2&\rho_is\alpha_i(x_{j0}+\rho_jc\alpha_j-x_{i0}-\rho_ic\alpha_i)\\
						-2&\rho_ic\alpha_i(y_{j0}+\rho_js\alpha_j-y_{i0}-\rho_is\alpha_i),\\
						\frac{dQ_{ij}}{d\beta_i}=~~0.&
					\end{aligned}\right.
					\label{Equ:DerivConstraintsC}
				\end{equation}
			\paragraph{Kinematic Constraints}
				Matrix~$\mathbf{S}$ is defined in appendix~\ref{Sec:ConstraintsMatrices} and represents the infinitesimal kinematic constraints such that~$\mathbf{S}d\boldsymbol{\theta}=\mathbf{0}, \forall d\boldsymbol{\theta}$. It is equal to~${d\boldsymbol{\mathcal{K}}}/{d\boldsymbol{\theta}}$. 
				Matrices~$\mathbf{S}_\psi$ and~$\mathbf{S}_\lambda$ are constructed using the corresponding columns of matrix~$\mathbf{S}$, namely
				\begin{equation}
					\mathbf{S}_\lambda=\left[\mathbf{S}_{\alpha_1};\mathbf{S}_{\beta_1};\dots;\mathbf{S}_{\beta_3}\right]
					\text{~and~}
					\mathbf{S}_\psi=\left[\mathbf{S}_{\rho_1};\mathbf{S}_{\rho_2};\mathbf{S}_{\rho_3}\right].
					\label{Equ:Schi}
				\end{equation}
		\subsubsection{Stiffness Matrix}
			The formulation presented in equation~(\ref{Equ:AlterCartStiff}) with the details given in appendix~\ref{Sec:AlterKchi} is used in this application, namely
			\begin{equation}
				\mathbf{K}_C=\mathbf{J}^{-T}\left[\mathbf{R}^T(\mathbf{K}_\theta+\mathbf{K}^\theta_E)\mathbf{R}+\mathbf{K}_R\right]\mathbf{J}^{-1}.
			\end{equation}
			\paragraph{Matrices~$\mathbf{G}$,~$\mathbf{R}$,~$\mathbf{J}_\lambda$ and~$\mathbf{J}$}
				Since all components of matrices~$\mathbf{S}_\psi$ and~$\mathbf{S}_\lambda$ are explicitly known, a formal expression of matrix~$\mathbf{G}$ could theoretically be obtained. However, the inversion of the~$6\times6$ matrix will lead to a very complex expression, and it is therefore simpler to compute~$\mathbf{G}$ numerically, i.e., compute~$\mathbf{S}_\psi$ and~$\mathbf{S}_\lambda$ from their formal expression and then  compute the inversion and multiplication as given in  appendix~\ref{Sec:StructureR}.

				Matrix~$\mathbf{J}_\lambda$ corresponds to the last 6 columns of matrix~$\mathbf{J}_\theta$ and matrix~$\mathbf{J}$ is obtained by right-multiplying matrix~$\mathbf{J}_{\theta}$ by matrix~$\mathbf{R}$~(eq.(\ref{Equ:JacobianGeneralized})). 
			\paragraph{Matrix~$\mathbf{K}^\theta_E$}
				By definition (eq.\ref{Equ:definitionKGapp}), matrix~$\mathbf{K}^\theta_E$ requires taking the derivative of~$\mathbf{J}_{\theta}$. This differentiation can be preferably performed formally  in order to avoid round-off errors due to a numerical derivation. Moreover, to avoid manipulating a tensor of~$3^{\textrm{rd}}$ order, matrix~$\mathbf{K}^\theta_E$ is calculated as
				\begin{equation}
					\mathbf{K}^\theta_E=\text{Jacobian}(\mathbf{J}_{\theta}^T\mathbf{f}, \boldsymbol{\theta})	
				\end{equation}
				where~$\mathbf{f}=\left[f_x;f_y;m_\phi\right]^T$ is considered constant.
			\paragraph{Matrix~$\mathbf{K}_R$}
				This matrix is more complicated to compute, because obtaining a formal expression of~$\mathbf{R}$ is almost impossible for the~3-RPR mechanism. Therefore, the alternative formulation of~$\mathbf{K}_R$ detailed in appendix~\ref{Sec:AlterKchi} is used. The algorithm used to implement and compute matrix ~$\mathbf{K}_R$ without introducing numerical errors, is presented below.
				\begin{itemize}
					\item[$\bullet$] Calculate formal expressions of  matrices~$\mathbf{S}_\lambda$ and~$\mathbf{S}_\psi$.
					\item[$\bullet$] Calculate a formal expression of matrices~$\mathbf{M}_\rho$ and~$\mathbf{M}_\lambda$, with the constant vector ($\overline{\mathbf{v}}=\left[v_1;\cdots;v_6\right]^T$) :
						\begin{equation}
							\begin{aligned}
								\medskip\mathbf{M}_\rho&=\text{Jacobian}(\mathbf{S}_\psi^{T}\overline{\mathbf{v}},\boldsymbol{\theta}),\\
								\mathbf{M}_\lambda&=\text{Jacobian}(\mathbf{S}_\lambda^{T}\overline{\mathbf{v}},\boldsymbol{\theta}).
							\end{aligned}
						\end{equation}
					\item[$\bullet$] Compute numerically vectors~$\mathbf{s}_\lambda$ and~$\mathbf{v}$ :
					\begin{equation}
						\mathbf{s}_\lambda=\mathbf{K}_\lambda(\boldsymbol{\lambda}-\boldsymbol{\lambda}_0)-\mathbf{J}_{\lambda}^T\mathbf{f}
						\text{~~and~~}
						\mathbf{v}=\mathbf{S}_\lambda^{-T}\mathbf{s}_\lambda.
					\end{equation}
					\item[$\bullet$] Assign the numerical value of the components of~$\mathbf{v}$ to variables~$v_i$ to enable the computation of~$\mathbf{M}_\lambda$ and~$\mathbf{M}_\rho$.
					\item[$\bullet$] Compute numerically~$\mathbf{R}$ (appendix~\ref{Sec:StructureR}).
					\item[$\bullet$] Finally compute~$\mathbf{K}_R$ :
						\begin{equation}
							\mathbf{K}_R=(-\mathbf{M}_\rho+\mathbf{S}_\psi^{-T}\mathbf{S}_\lambda^{-T}\mathbf{M}_\lambda)\mathbf{R}.
						\end{equation}
				\end{itemize}
			\paragraph{Stiffness Matrix}
				With the above matrices and vectors, the CSM of the 3-RPR mechanism can be computed in any non-singular configuration, using the formulation presented in eq.(\ref{Equ:AlterCartStiff}).
	\subsection{Simulation of the Mechanism}
		In order to illustrate the validity and the accuracy of the proposed SM, a simple application is presented below.

		The trajectory followed by the~3-RPR mechanism subjected to an external wrench (applied on its end-effector) is computed using the SM. Actually, each increment of the external wrench multiplied by the SM computed in the local configuration provides an incremental displacement and the combination of all these displacements enables to plot the trajectory. In other words, the trajectory is computed with the following expression, implemented numerically:
		\begin{equation}
 			\mathbf{x}\leftarrow\mathbf{x}+\mathbf{K}_C^{-1}\delta\mathbf{f}.
		\end{equation}
		On the other hand, the results of the commercial software MSC.~Adams are used as references to evaluate the accuracy of the computations and indirectly to prove the validity of the presented SM. In MSC.~Adams, the equilibrium and the position of the mechanism are computed at each step and therefore there is no drift due to an iterative method. Moreover, by choosing the static simulation option, the dynamical effects are not taken into account, which is consistent with our assumptions. 
		The wrench applied on the reference point of the platform is 
			\begin{equation}
				\mathbf{f}(t)=\left[f_0\sin(2\pi t); f_0\sin(4\pi t); 0\right]^T,~f_0=100N.
				\label{Equ:Wrenchft}
			\end{equation}
		\subsubsection{Comparison with Other Formulations}
			In this subsection, we consider a mechanism in which the stiffness of the actuators are finite but the stiffness of the passive joints is equal to zero. 

			With this simulation, we can compare~($a$) the accuracy of the SM presented by Salisbury~\cite{Salisbury}, ($b$) the accuracy of the SM presented by Chen and Kao~\cite{CCT1} and~($c$) the accuracy of the proposed SM. The SM for PMs presented in this paper is noted PM~($c$). Matrices~($c$) and~($b$) represents the conservative congruence transformation (CCT).

\begin{figure}[t]
					\centering
					\includegraphics[height=5cm]{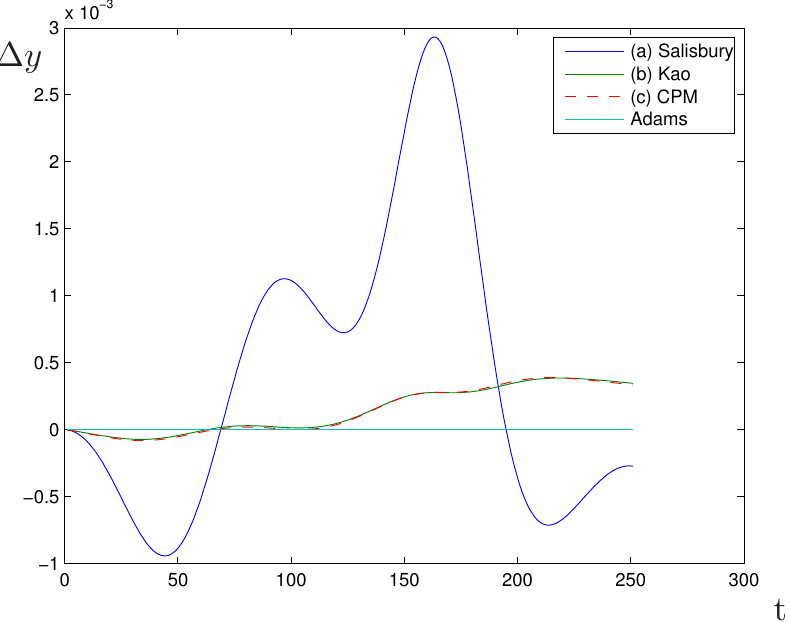}
					\caption{Discrepancy in~$x$-coordinate of the pose of the mechanism subjected to~$\mathbf{f}(t)$.}
					\label{Fig:Xconventionnal}
				\end{figure}

\begin{figure}[t]
					\centering
					\includegraphics[height=5cm]{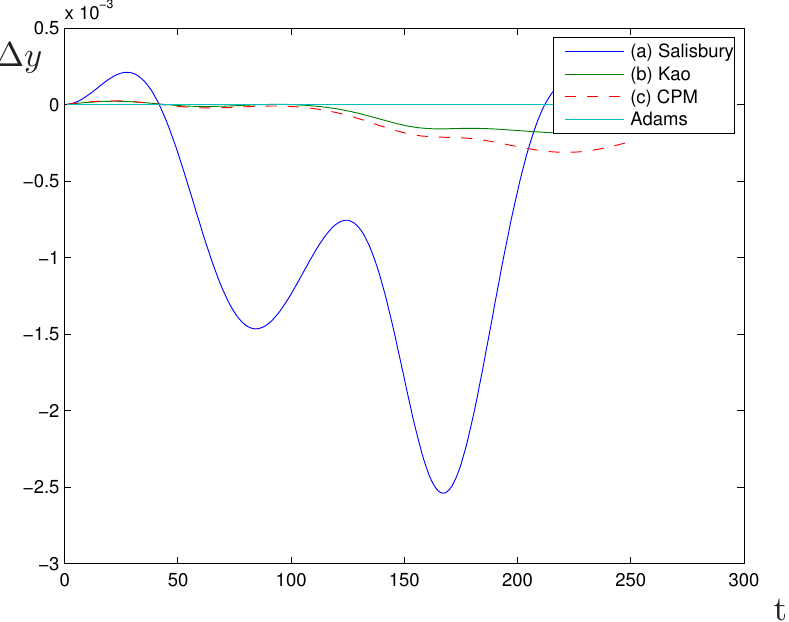}
					\caption{Discrepancy in~$y$-coordinate of the pose of the mechanism subjected to~$\mathbf{f}(t)$.}
					\label{Fig:Yconventionnal}
				\end{figure}

\begin{figure}[t]
					\centering
					\includegraphics[height=5cm]{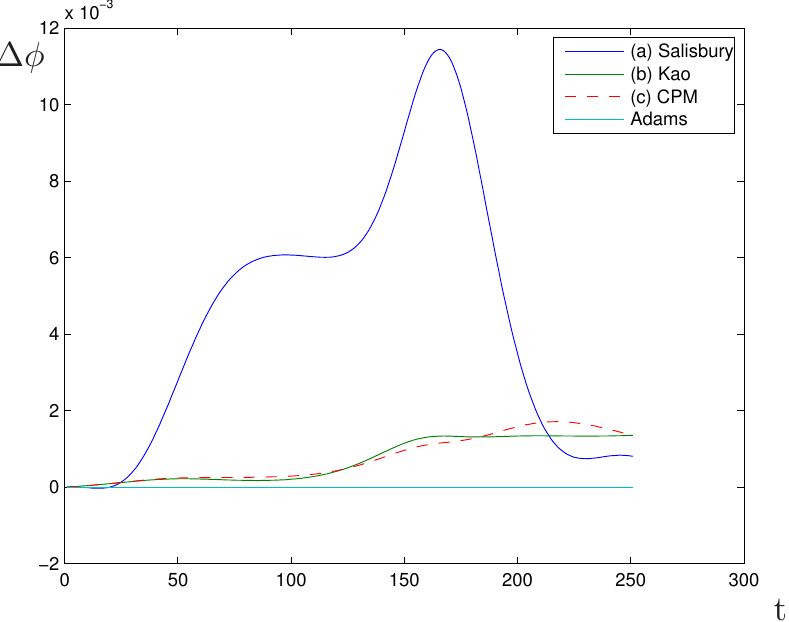}
					\caption{Discrepancy in~$\phi$-coordinate of the pose of the mechanism subjected to~~$\mathbf{f}(t)$.}
					\label{Fig:Phiconventionnal}
				\end{figure}

			To make the comparison between all these matrices, matrix~($c$) is computed with the value of stiffness of the passive joints equal to zero ($\mathbf{K}_{\lambda}=\mathbf{0}$). Thus, formulations~($b$) and~($c$) become equivalent, but since their implementation are different, mainly because of the alternative formulation used here, their computation can provide slightly different results.
			The three expressions of~$\mathbf{K}_C$ are written as
			\begin{equation}
				\begin{aligned}
					\medskip\text{($a$)~\begin{footnotesize}Salisbury\end{footnotesize}}:&~\mathbf{K}_C=\mathbf{J}_{\rho}^{-T}\mathbf{K}_\rho\mathbf{J}_{\rho}^{-1}\\
					\medskip\text{($b$) Chen}~~\,:&~\mathbf{K}_C=\mathbf{J}_{\rho}^{-T}\left(\mathbf{K}_\rho+\mathbf{K}^\rho_E\right)\mathbf{J}_{\rho}^{-1}\\
					\text{($c$) PM}~~~~\,:&~\mathbf{K}_C=\mathbf{J}^{-T}\left(\mathbf{K}_\psi+\mathbf{K}_I+\mathbf{K}_E\right)\mathbf{J}^{-1}
				\end{aligned}
			\end{equation}
			where~$\mathbf{J}_{\rho}$ is the Jacobian matrix usually used for the~3-RPR mechanism,~$\mathbf{K}_\rho$ is the~$3\times3$ diagonal matrix representing the stiffness of the actuators~$\rho_i$ and~$\mathbf{K}^\rho_E$ corresponds to the matrix~$(-\mathbf{K}_G)$ defined in \cite{CCT1}. In cases~($a$) and~($c$), the coordinates of the passive joints~$\alpha_i$ and~$\beta_i$ do not appear. The time of simulation~$t$ varies from~$0$\,s to~$1$\,s in~250~iterations, such that~$\delta t=1/250$\,s (4\,ms). The increment of external wrench is~$\delta \mathbf{f}=\mathbf{f}'(t)\delta t$. And the stiffness values of the joints used in this section are~$k_\alpha=k_\beta=0$\,N.rad$^-1$ and~$k_\rho=2000$\,N.mm$^-1$.
			\paragraph{Results}
				\begin{figure}[t]
					\centering
					\includegraphics[height=5cm]{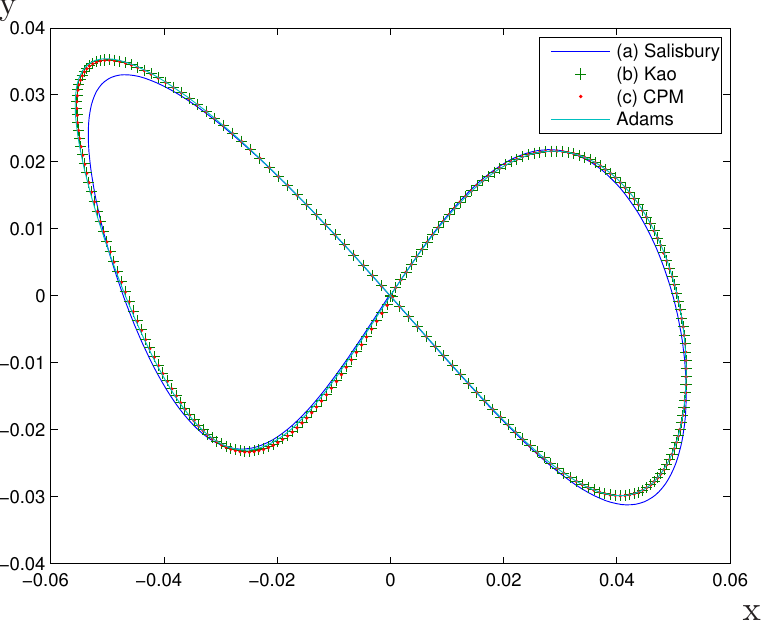}
					\caption{Trajectory~$(x,y)$ described by the mechanism subjected to~$\mathbf{f}(t)$.}
					\label{Fig:Trajconventionnal}
				\end{figure}
				Figure~\ref{Fig:Trajconventionnal} shows the trajectory described by the mechanism, computed with the software MSC.~Adams and with the four matrices. It can be noticed that the results obtained with matrix ($a$) does not correspond to the trajectory computed with MSC.~Adams, while results of the CCT matrices ($b$) and ($c$) are accurate. Since the results are very close to each other, they are presented in another form in Figs.~\ref{Fig:Xconventionnal}, \ref{Fig:Yconventionnal} and \ref{Fig:Phiconventionnal}. The latter graphs show the difference in the~3 components ($x$,$y$,$\phi$) of the pose~$\mathbf{x}$, between the reference from MSC.~Adams and the computation with each matrix.

				Some important points can be noted on the graphs:\\
				$\bullet$~The discrepancy between the results obtained with MSC.~Adams and with the CCT matrices increases uniformly as the simulation proceeds. This effect is a drift due to the iterative computation. In a real use of these matrices, the variables are updated by a measurement on the robot at each step, thus this drift should disappear.\\
 				$\bullet$~On the contrary, the discrepancy between the results obtained with MSC.~Adams and Salisbury's matrix is clearly a function of the external loads. The larger these loads are, the larger the error in the stiffness computation will be. The drift due to the iterative computation also exists but it is secondary compared to the effect of loads. We can however observe that the error due to the load seems to be compensated for since the error decreases when the load decreases.\\
 				$\bullet$~As shown in Fig.~\ref{Fig:Trajconventionnal}, the simulation with the CCT matrices are much more accurate than with Salisbury's matrix. The range of deviation for the CCT matrices after 250 iterations is~$0,5\,\mu$m in position and~$2.10^{-3}$\,rad in orientation, while the maximal deviation for Salisbury's matrix is~$3\,\mu$m in position and~$1,2.10^{-2}$\,rad in orientation.\\
				$\bullet$~A small difference between matrices ($b$) and ($c$) can be noticed at the end of the simulation (notably in Fig.~\ref{Fig:Yconventionnal} and Fig.~\ref{Fig:Phiconventionnal}). These differences are only due to the numerical computations.
		\paragraph{Conclusion}
				This simulation confirms the validity and the equivalence of both CCT formulations. It also shows their accuracy. On the other hand, this simulation proves the invalidity of Salisbury's matrix. Indeed, if the latter matrix can seem acceptable for very small external wrenches such as vibrations, the error grows quickly with the loads.
		\subsubsection{Impact of the Passive Joints}
\begin{figure}[!tp]
					\centering
					\includegraphics[height=5cm]{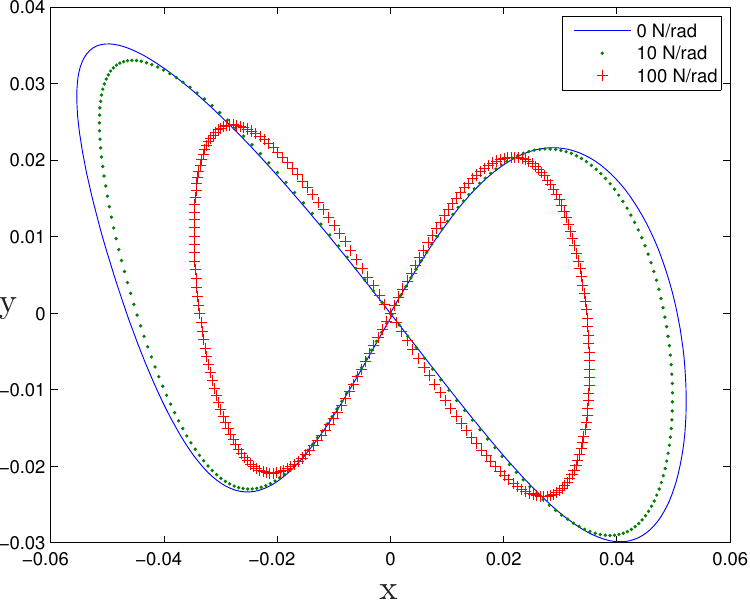}
					\caption{Trajectory~$(x,y)$ described by different mechanisms subjected to~$\mathbf{f}(t)$.}
					\label{Fig:Trajcompliant}
				\end{figure}

			The main novelty of the SM proposed in this paper is the possibility to take into account the stiffness of the passive joints. Figure~\ref{Fig:Trajcompliant} illustrates this new possibility and shows the trajectory performed by the mechanism when it is subjected to the external wrench~$\mathbf{f}(t)$ defined in eq.(\ref{Equ:Wrenchft}). This trajectory is computed with passive joints having different stiffness~$k_{\alpha_i}$ and~$k_{\beta_i}$, namely: 0\,N.rad$^{-1}$, 10\,N.rad$^{-1}$ and 100\,N.rad$^{-1}$. Note that the trajectories calculated with MSC.~Adams are not represented in figure~\ref{Fig:Trajcompliant} because they coincide exactly with those of our model. The discrepancy cannot be observed at this scale.

			As expected, it can be observed that stiffer passive joints give a stiffer mechanism and decrease the amplitude of the displacement due to external wrenches. The shape of the displacement is also affected. The curve drawn with crosses represents the trajectory of a mechanism in which the stiffness of the passive joint is only 20 times smaller than that of the actuators, but even in this extreme and almost unrealistic case, the trajectory is computed with a very good accuracy.
			
			The important point illustrated by this application is the possibility to accurately determine the behaviour of a mechanism subjected to external loads. Indeed, the stiffness of the passive joints can be regarded as an advantage or as a disadvantage in the context of control of a manipulator since it makes the manipulator less sensitive to external perturbations but requires more powerful actuators. However, if high precision is required, with the new SM that enables to compute accurately their behaviour, compliant joints with zero mechanical clearance offer only advantages.
			In other words, with the knowledge of the SM, the \emph{precision} of a mechanism becomes independent from its \emph{stiffness}.

\section{Conclusion}
	The proposed formulation of the stiffness matrix is clear and meaningful. The presented Cartesian stiffness matrix is a generalization of the already existing matrices published in the literature, since it can take into account non-zero external loads, non-constant Jacobian matrices, stiff passive joints and additional compliances, these two latter points being its main novelty. 

	Moreover, the results predicted with this stiffness matrix are very accurate and the proposed SM enables a very accurate control of parallel manipulators built with elastic joints.
\section*{Acknowledgements}
	The authors would like to acknowledge the financial support of the Natural Sciences and Engineering Research Council of Canada (NSERC) as well as the Canada Research Chair (CRC) Program.	

\bibliographystyle{asme_journals}

\bibliography{StiffnessMatrixJMR}

\appendix
\section{Matrices of Constraints}
	\subsection{Matrices~$\mathbf{S}$,~$\mathbf{S}_\lambda$ and~$\mathbf{S}_\psi$}
		\label{Sec:ConstraintsMatrices}
		From the geometric constraints (eq.(\ref{Equ:Contr_GeoGeneral})), the kinematic constraints of a PM can also be written as
		\begin{equation}
			d\boldsymbol{\mathcal{K}}(\boldsymbol{\theta})=\frac{d\boldsymbol{\mathcal{K}}(\boldsymbol{\theta})}{d\boldsymbol{\theta}}d\boldsymbol{\theta}=\mathbf{S}d\boldsymbol{\theta} = \mathbf{0},
			\label{Equ:MatrixS}
		\end{equation}
		where matrix is~$\mathbf{S}$ is defined as the derivative of~$\boldsymbol{\mathcal{K}}$ with respect to~$\boldsymbol{\theta}$. 
		Making the distinction between the generalized and the dependent coordinates, one can write
		\begin{equation}
			\mathbf{S}d\boldsymbol{\theta} =\left[\mathbf{S}_\psi;\mathbf{S}_\lambda\right]\begin{bmatrix}d\boldsymbol{\psi}\\d\boldsymbol{\lambda}\end{bmatrix}=\mathbf{S}_\psi d\boldsymbol{\psi}+ \mathbf{S}_\lambda d\boldsymbol{\lambda} = \mathbf{0},
			\label{Equ:SChiSLambda}
		\end{equation}
		where~$\mathbf{S}_\psi$ is the~$\mathfrak{C}\times \mathfrak{M}$ matrix composed of the~$\mathfrak{M}$ columns of~$\mathbf{S}$ corresponding to~$\boldsymbol{\psi}$ and ~$\mathbf{S}_\lambda$ is the~$\mathfrak{C}\times \mathfrak{C}$ matrix composed of the~$\mathfrak{C}$ columns of~$\mathbf{S}$ corresponding to~$\boldsymbol{\lambda}$.
	\subsection{Matrices  \textbf{G} and \textbf{R}} 
		\label{Sec:StructureR}
		By definition, the~$\mathfrak{C}$ coordinates~$\lambda_i$ are the solutions of the~$\mathfrak{C}$ geometrical constraints~$\boldsymbol{\mathcal{K}}$ for a set~$\boldsymbol{\psi}$, so~$\mathbf{S}_\lambda=d\boldsymbol{\mathcal{K}}/d\boldsymbol{\lambda}$ is a matrix of full rank and therefore is always invertible. Equation~(\ref{Equ:SChiSLambda}) is equivalent to
		\begin{equation}
				d\boldsymbol{\lambda}=-\mathbf{S}_\lambda^{-1}\mathbf{S}_\psi d\boldsymbol{\psi}.
			\label{Equ:dLambda}
		\end{equation}
		Thus, matrices~$\mathbf{G}$ and~$\mathbf{R}$ are expressed as 
		\begin{equation}
			\mathbf{G}=-\mathbf{S}_\lambda^{-1}\mathbf{S}_\psi
			\text{~~and~~}
			\mathbf{R}=\begin{bmatrix}\mathbf{1}_\mathfrak{M}\\-\mathbf{S}_\lambda^{-1}\mathbf{S}_\psi\end{bmatrix}.
			\label{Equ:StructureR}
		\end{equation}
\section{Implementation of the Stiffness Matrix}
	\label{Sec:AlterKchi}
	\subsection{Alternative Formulation of~$\mathbf{K}_C$}
		\subsubsection{Matrix~$\mathbf{K}_E$}
			\label{Sec:AlterKE}
			In some PMs, a formal expression of~$\mathbf{J}$ as a function of~$\boldsymbol{\psi}$ can be difficult to obtain whereas~$\mathbf{J}_\theta$ is easy to formulate as a function of coordinates~$\theta_i$. And~$\mathbf{G}$ and~$\mathbf{R}$ are defined as function of~$\boldsymbol{\theta}$. Therefore, it is generally more interesting to calculate~$d(\cdot)/d\boldsymbol{\theta}$ instead of~$d(\cdot)/d\boldsymbol{\psi}$. 
			Using equation~(\ref{Equ:JacobianGeneralized}), the definition of~$\mathbf{K}_E$ (eq.(\ref{Equ:MatKE})) is equivalent to
			\begin{equation}
				\begin{aligned}
					\mathbf{K}_E=&-\frac{d\mathbf{J}}{d\boldsymbol{\psi}}^T\mathbf{f}=-\frac{d(\mathbf{J}_\theta\mathbf{R})}{d\boldsymbol{\psi}}^T\mathbf{f}\\&=-\mathbf{R}^T(\frac{d\mathbf{J}_\theta}{d\boldsymbol{\theta}}^T\mathbf{f})\frac{d\boldsymbol{\theta}}{d\boldsymbol{\psi}}-\frac{d\mathbf{R}}{d\boldsymbol{\psi}}^T\mathbf{J}_\theta^T\mathbf{f}
				\end{aligned}
				\label{Equ:KEintermediaire}
			\end{equation}
			In this equation, matrix~$(-d\mathbf{J}_\theta/d\boldsymbol{\theta})^T\mathbf{f}$  is noted~$\mathbf{K}^\theta_E$. 
			In this matrix, the kinematic constraints are not taken into account, each leg is considered as an independent mechanism. 
			Hence, since~$\mathbf{R}^T=\left[\mathbf{1};\mathbf{G}^T\right]$, the last term of eq.(\ref{Equ:KEintermediaire}) is noted~$\mathbf{K}_{EG}$ and can be calculated as
			\begin{equation}
				\begin{aligned}
					\mathbf{K}_{EG}=-\frac{d\mathbf{R}}{d\boldsymbol{\psi}}^T\mathbf{J}_\theta^T\mathbf{f}&=-\frac{d\mathbf{1}}{d\boldsymbol{\psi}}^T\mathbf{J}_\psi^T\mathbf{f}-\frac{d\mathbf{G}}{d\boldsymbol{\psi}}^T\mathbf{J}_\lambda^T\mathbf{f}\\
					&=-\left[\frac{d\mathbf{G}}{d\boldsymbol{\theta}}^T\mathbf{J}_\lambda^T\mathbf{f}\right]\mathbf{R}.
				\end{aligned}
				\label{Equ:KEintermediaire2}
			\end{equation}
			where~$\mathbf{J}_\lambda=d\mathbf{x}_c/d\boldsymbol{\lambda}$ contains the columns of~$\mathbf{J}_\theta$ corresponding to coordinates~$\lambda_i$. Thus eq.(\ref{Equ:KEintermediaire}) can be written~as
			\begin{equation}
				\mathbf{K}_E=\mathbf{R}^T\mathbf{K}^\theta_E\mathbf{R}+\mathbf{K}_{EG}.
				\label{Equ:definitionKGapp}
			\end{equation}
		\subsubsection{Matrix~$\mathbf{K}_I$}
			\label{Sec:MatKS}
			A matrix~$\mathbf{K}_R$ that represents the effects of the change of the constraints~$\mathbf{G}^T$ is defined by
			\begin{equation}
				\mathbf{K}_R=\mathbf{K}_{IG}+\mathbf{K}_{EG}=\frac{d\mathbf{G}}{d\boldsymbol{\psi}}^T\boldsymbol{\tau}_\lambda-\frac{d\mathbf{G}}{d\boldsymbol{\psi}}^T\mathbf{J}_\lambda^T\mathbf{f}.
			\end{equation}
			Thus, with the same operations as in section (\ref{Sec:AlterKE}),~$\mathbf{K}_R$ is calculated as 
			\begin{equation}
				\mathbf{K}_R=\left[\frac{d\mathbf{G}}{d\boldsymbol{\theta}}^T(\boldsymbol{\tau}_\lambda-\mathbf{J}_\lambda^T\mathbf{f})\right]\mathbf{R}=\left[\frac{d\mathbf{G}}{d\boldsymbol{\theta}}^T\mathbf{s}_\lambda\right]\mathbf{R}.
				\label{Equ:KS}
			\end{equation}
			where~$\mathbf{s}_\lambda$ represents the sum of the forces/torques applied on the constrained joints.
		\subsubsection{Cartesian Stiffness Matrix}
			Using the above matrices,~$\mathbf{K}_C$ can be written as
			\begin{equation}
				\mathbf{K}_C=\mathbf{J}^{-T}\left(\mathbf{K}_\psi+\mathbf{G^T}\mathbf{K}_\lambda\mathbf{G}+\mathbf{K}_R+\mathbf{R}^T\mathbf{K}^\theta_E\mathbf{R}\right)\mathbf{J}^{-1}.
			\end{equation}
			This latter equation being equivalent to
			\begin{equation}
				\mathbf{K}_C=\mathbf{J}^{-T}\left[\mathbf{R}^T(\mathbf{K}_\theta+\mathbf{K}^\theta_E)\mathbf{R}+\mathbf{K}_R\right]\mathbf{J}^{-1}
				\label{Equ:AlterCartStiff}
			\end{equation}
	\subsection{Computation of Matrix~$\mathbf{K}_R$}
		\label{Sec:AlterKR}
		Matrix~$\mathbf{K}_R$ results from the differentiation of matrix~$\mathbf{G}$. But since a formal expression of~$\mathbf{G}$ might be too complex to be handled in closed form due to the inversion of the~$(\mathfrak{C}\times \mathfrak{C})$ matrix~$\mathbf{S}_\lambda$, an alternative method to calculate it can be used. Moreover, the derivative of a matrix with respect to a vector gives a tensor of~$3^{\textrm{rd}}$ order, and this type of mathematical object and all its associated functions are usually not developed in most current software packages.
		For the above reasons, a detailed formulation of~$\mathbf{K}_R$ is presented below. This formulation is easier to implement and it enables the computation of~$\mathbf{K}_R$ without introducing numerical inaccuracies.
	
		To avoid taking the derivative of matrix~$\mathbf{S}$ with respect to vector~$\boldsymbol{\theta}$ ---which gives a tensor of~$3^{\textrm{rd}}$ order--- the vector~$\mathbf{G}^T\mathbf{s}_\lambda$ is first calculated 
		\begin{equation}
			\mathbf{G}^T\mathbf{s}_\lambda=-\mathbf{S}_\lambda^{-1}\mathbf{S}_\psi^T\mathbf{s}_\lambda.
			\label{Equ:RtSappl}
		\end{equation}
		Then, the Jacobian matrix of~$\mathbf{G}^T\mathbf{s}_\lambda$ with respect to~$\boldsymbol{\theta}$ can be calculated, considering~$\mathbf{s}_\lambda$ as a constant vector (noted~$\overline{\mathbf{s}_\lambda}$) 
		\begin{equation}
			\frac{d\mathbf{G}^T}{d\boldsymbol{\theta}}\mathbf{s}_\lambda=\frac{d(\mathbf{G}^T\overline{\mathbf{s}_\lambda})}{d\boldsymbol{\theta}}=\frac{d(-\mathbf{S}_\psi^T\mathbf{S}_\lambda^{-T})}{d\boldsymbol{\theta}}\overline{\mathbf{s}_\lambda}.
			\label{Equ:dRtSappl1}
		\end{equation}
		Equation (\ref{Equ:dRtSappl1}) is equivalent to :
		\begin{equation}
			\frac{d\mathbf{G}^T}{d\boldsymbol{\theta}}\mathbf{s}_\lambda=-\frac{d\mathbf{S}_\psi^{T}}{d\boldsymbol{\theta}}\mathbf{S}_\lambda^{-T}\overline{\mathbf{s}_\lambda}-\mathbf{S}_\psi^{T}\frac{d\mathbf{S}_\lambda^{-T}}{d\boldsymbol{\theta}}\overline{\mathbf{s}_\lambda}.
			\label{Equ:dRtSappl2}
		\end{equation}
		Since calculating a formal expression of~$\mathbf{S}_\lambda^{-1}$ might be too complex and since a formal derivative is desired to avoid any round-off errors in the computation of~$\mathbf{K}_S$, the following equivalent formulation of~$\mathbf{S}_\lambda^{-T}$  is used. In this equation, the inversion can be computed numerically but the derivative can be obtained formally.
		\begin{equation}
			\frac{d\mathbf{S}_\lambda^{-T}}{d\boldsymbol{\theta}}=-\mathbf{S}_\lambda^{-T}\frac{d\mathbf{S}_\lambda^T}{d\boldsymbol{\theta}}\mathbf{S}_\lambda^{-T}.
			\label{Equ:dmatrixinverse}
		\end{equation}
		Thus, equation (\ref{Equ:dRtSappl2}) is equivalent to
		\begin{equation}
			\frac{d\mathbf{G}^T}{d\boldsymbol{\theta}}\mathbf{s}=-\frac{d\mathbf{S}_\psi^{T}}{d\boldsymbol{\theta}}\mathbf{S}_\lambda^{-T}\overline{\mathbf{s}_\lambda}+\mathbf{S}_\psi^{T}\mathbf{S}_\lambda^{-T}\frac{d\mathbf{S}_\lambda^T}{d\boldsymbol{\theta}}\mathbf{S}_\lambda^{-T}\overline{\mathbf{s}_\lambda}.
			\label{Equ:dRtSappl3}
		\end{equation}
		Here again, the derivative of a matrix with respect to a vector is required. To avoid such a derivative, the following vectors are introduced :
		\begin{equation}
			\left\{\begin{aligned}
				\mathbf{v}&=\mathbf{S}_\lambda^{-T}\mathbf{s}_\lambda\\
				\mathbf{m}_\psi&=\mathbf{S}_\psi^T\mathbf{S}_\lambda^{-T}\mathbf{s}_\lambda=\mathbf{S}_\psi^T\underline{\mathbf{v}}\\
				\mathbf{m}_\lambda&=\mathbf{S}_\lambda^T\mathbf{S}_\lambda^{-T}\mathbf{s}_\lambda=\mathbf{S}_\lambda^T\underline{\mathbf{v}}\\
			\end{aligned}\right.
			\label{Equ:vectorVmm}
		\end{equation}
		For practical purposes, the vectors~$\mathbf{m}_\psi$ and~$\mathbf{m}_\lambda$ are formally calculated with a vector of constant components  ($\underline{\mathbf{v}}=\left[v_1,\cdots,v_\mathfrak{C}\right]^T$), then the formal derivatives are calculated. Finally, the numerically computed components of~$\mathbf{v}$ are assigned to variables~$v_i$ to obtain matrices~$\mathbf{M}_\psi$ and~$\mathbf{M}_\lambda$.
		\begin{equation}
			\left\{\begin{aligned}
				\medskip\frac{d\mathbf{S}_\psi^{T}}{d\boldsymbol{\theta}}\mathbf{S}_\lambda^{-T}\mathbf{s}_\lambda&=\frac{d\mathbf{m}_\psi}{d\boldsymbol{\theta}}=\mathbf{M}_\psi\\
				\frac{d\mathbf{S}_\lambda^{T}}{d\boldsymbol{\theta}}\mathbf{S}_\lambda^{-T}\mathbf{s}_\lambda&=\frac{d\mathbf{m}_\lambda}{d\boldsymbol{\theta}}=\mathbf{M}_\lambda\\
			\end{aligned}\right.
			\label{Equ:vectorVmm2}
		\end{equation}	
		An alternative formulation of matrix~$\mathbf{K}_R$ defined in eq.(\ref{Equ:KS}) can then be written as 
		\begin{equation}
			\mathbf{K}_R=\left[-\frac{d\mathbf{S}_\psi^{T}}{d\boldsymbol{\theta}}\mathbf{S}_\lambda^{-T}\mathbf{s}_\lambda+\mathbf{S}_\psi^{T}\mathbf{S}_\lambda^{-T}\frac{d\mathbf{S}_\lambda^T}{d\boldsymbol{\theta}}\mathbf{S}_\lambda^{-T}\mathbf{s}_\lambda\right]\mathbf{R}.
			\label{Equ:ComputKR}
		\end{equation}

%
%
%
%
%
%
%

\end{document}